# A Real-Time Approach for Smart Building Operations Prediction Using Rule-Based Complex Event Processing and SPARQL Query

Shashi Shekhar Kumar[a*]

Department of Information Technology

Indian Institute of Information Technology Allahabad, Prayagraj, India

Ritesh Chandra[b]

Department of Information Technology

Indian Institute of Information Technology, Allahabad, Prayagraj, India

Sonali Agarwal[c]

Department of Information Technology

Indian Institute of Information Technology, Allahabad, Prayagraj, India

**Abstract**

Due to intelligent, adaptive nature towards various operations and their ability to provide maximum comfort to the occupants residing in them, smart buildings are becoming a pioneering area of research. Since these architectures leverage the Internet of Things (IoT), there is a need for monitoring different operations (Occupancy, Humidity, Temperature, $CO_2$, etc.) to provide sustainable comfort to the occupants. This paper proposes a novel approach for intelligent building operations monitoring using rule-based complex event processing and query-based approaches for dynamically monitoring the different operations. Siddhi is a complex event processing engine designed for handling multiple sources of event data in real time and processing it according to predefined rules using a decision tree. Since streaming data is dynamic in nature, to keep track of different operations, we have converted the IoT data into an RDF dataset. The RDF dataset is ingested to Apache Kafka for streaming purposes and for stored data we have used the GraphDB tool that extracts information with the help of SPARQL query. Consequently, the proposed approach is also evaluated by deploying the large number of events through the Siddhi CEP engine and how efficiently they are processed in terms of time. Apart from that, a risk estimation scenario is also designed to generate alerts for end users in case any of the smart building operations need immediate attention. The output is visualized and monitored for the end user through a tableau dashboard.

Keywords- Complex Event Processing, Rule based parameters, SPARQL, Event driven decision, GraphDB, Decision tree.

# 1 Introduction

By the year 2050, almost all common devices will be online, according to a recent survey conducted by Ericsson [1]. These IoT-enabled devices collect data from physically deployed sensors and constantly collect and react to that data, and these vast levels of data can be unlocked for further unlocking useful information. Occupant comfort is an important prospect in the scenarios of smart buildings, and it needs to be monitored constantly so that if any parameter goes up or down or exceeds the predefined values, an alert may be generated at the appropriate time and the decision may be taken in a dynamic environment. The rules for different parameters ($CO_2$, temperature, occupancy, heating, ventilation, etc.) are defined using a decision tree, which classifies the data based on labeled values.

A smart building system that uses rules to determine behavior and automation is known as a rule-based smart building system. These regulations specify the circumstances and events under which the building's systems and equipment must operate. Apart from that, a powerful open-source platform for real-time data processing and event-driven applications is Siddhi CEP (Complex Event Processing). Siddhi CEP may be used alongside other technologies to develop smart building functions and even if it can process real time events also.

Rule-based Complex Event Processing (CEP) is a method for detecting and responding to complex events in real-time by fusing rule-based systems with event processing tools. When using rule-based CEP, events are processed and assessed in accordance with a set of standards to identify relevant patterns and initiate the necessary actions.

Consequently, the objective of our work is divided into two phases. In the first phase of the work, we demonstrate real-time streaming processing through the apache kafka tool and Apache Flink event stream processing. In the second phase, we process the event streams using the query processing through SPARQL, which can query the RDF dataset to find what is going through the stream data. The rules are formed using a decision tree. Existing smart building operations [2] are based on manual implementation of rules, which are not as efficient for monitoring smart building operations in real-time contexts.

In this paper, we present a novel approach for smart building operations monitoring using a complex event processing engine. It is based on methods of complex event processing combined with rule-based terminology for fetching the real-time results using query-based languages. The current approach collects (1) various smart building operation parameters (2) processed through Kafka event streaming (3) uses Flink for CEP implementation; (4) Use of siddhi for events deployment  and (5) queries performed on the data streams with GraphDB tool. The novel part of this work is that a rule-based approach to performing a query will help smart building occupants make effective decisions in terms of smart building operations. We evaluate the proposed model using various accuracy measures and query processing times to consolidate this work.

The rest of this paper is organised as follows: Section **II** presents the existing approaches for intelligent building operations proposed in recent times. Section **III** discusses the proposed architecture, including its components and workings. Section **IV** elaborates the experiment details and results. Section **V** concludes the work and outlines the future directions.

## 2 Related Work

In this section, we discuss the related work on smart building operations and existing approaches with respect to rule-based implementation. Later we discussed various works related to complex event processing for intelligent building operations. Additionally we relate the approaches that work with real-time data and query based processing for SPARQL.CEP with RDF dataset is also reviewed in the later section.

### 2.1 Building Operations prediction system

Dong et al. [3] proposed a pattern-based algorithm approach for maintaining the occupants comforts and behaviors while reducing their energy consumption by 30%. The results of the proposed approach were evaluated using training and testing using energy simulation tools, which predicted better results as compared to other HVAC control strategies.

Mofidi et al. [4] proposed a well-structured architecture for computational intelligence and optimal decision-making for indoor occupants inside the buildings. Their work was inclined towards various parameters, including occupancy modeling behavior, building control, and optimizations.

Khan et al. [5] introduced a novel approach using a hybrid AI framework for accurate forecasting using power consumption and generation. The entire work was carried out using three steps, which include (i) data refinement procedures and (ii) in the next phase, the refined data is used for convolutional long short term memory (ConvLSTM) for discriminative pattern learning using previous historical data and a bidirectional gated recurrent unit for extracting the temporal aspects. After forecasting the model is evaluated using statistical error mechanism for effectiveness of the predictions.

Kim et al. [6] proposed a machine-learning-based model for estimating the space occupied using $CO_2$ concentration and ventilation system operation status. The model was trained using ANN (Artificial Neural Network) and RFC (random forest classifier) that measured an accuracy of 0.9102 and 0.9180 %. Apart when $CO_2$ concentration and differential pressure data were used the accuracy was reduced significantly to 0.8916% and 0.8936%.

### 2.2 Intelligent Building in Context of Big data

Chen et al. [7] explored an area of further research for BIM (Building Information Modelling) and IoT integration of smart building transformation processes. The integration highlights sustainable building development, which needs to be assessed with respect to human and social dimensions. BIM provides a platform for connecting the stakeholders involved for a lifetime, which allows occupants residing inside to make effective decisions. Further, they added big data and cloud computing for leveraging operations.

Peng et al. [8] emphasized the importance of IoT and its integration with big data for developing sustainable smart building systems. The integration can track and monitor in real time, which can enhance the IoT system's adaptability. They also highlighted the complex risks associated with various crowd gatherings and fire incidents, which occur frequently.

Zand et al. [9] divided the smart building work into two folds: artificial intelligence methods and classification mathematical models. The essential part of their work covers rules based on intelligent smart homes and challenges related to deployment to form a prototype using sensors and actuators. They also highlighted the protocols and standards needed to design smart buildings.

Pinto et al. [10] proposed a comprehensive review from the perspective of big data and transfer learning for smart building operations. Their approach shows a depth of analysis of recent work with algorithms, metrics, and applications. Their review was inclined to focus on four areas of applications, which include: (1) load prediction inside smart buildings (2) Occupancy detection activities (3) smart building dynamics; (4) smart energy monitoring.

Kumar et al. [11] proposed a novel approach for smart energy consumption tracking and alerting using big data and IoT systems.The model can track the energy usage and remaining available units in real-time. From the big data perspective, the model can efficiently track usage and help in making effective decisions using predictive machine learning approaches.

Khanpara et al. [12] proposed a security-based context-aware scheme for IoT-driven smart homes to protect against security-based threats. Their work explores in detail the various threats that can pose challenges to the indoor occupants inside the smart building premises. The result shows that the approach is effective in terms of cost, performance, and maintenance parameters.

Alexandru et al. [13] proposed a thermal analysis model to control heat transfer for residential purposes using lumped parameters. The parameters are generalized further using long short-term memory (LSTM) neural networks. The result shown using the model is quite promising in terms of the efficiency of the temperature controller for overshoot and energy consumption.

## 2.3 IoT driven complex event processing

Brazález et al. [14] proposed a FUME-based decision support system monitoring using AQ standards for reducing the air pollutants. The model uses fuzzy logic-based complex event processing for maintaining the parameters and improving decision-making. The model was used for a particular case study, and the result shows that it violates the world health parameters.

Roldán-Gómez et al. [15] introduced an approach to generate complex event processing-based rules for real-time attack detection in an unsupervised manner. They integrated the CEP model using various algorithms such as gaussian mixture models (GMM), principal component analysis (PCA), and Mahalanobis distance.The architecture was tested using various experiments using rule generation and achieved an F1 score of 0.9890% accuracy.

Verma et al. [16] proposed a system called RACER for end-to-end complex event processing of varying changes and observations in sensor-generated data. The result shows that using the proposed approach is much better as compared to other state-of-the-art methodologies.

**2.4 RDF with Kafka and Flink.**

Ren et al. proposed a design for RDF stream processing [17] through big data frameworks Apache Kafka and Apache Spark that is scalable,fault-tolerant, and high-performing for stream events. Their emphasis on implementation was to ease the complicated architecture by leveraging various libraries of frameworks.

Ren et al. [18] introduced a novel approach of a stride-based hybrid RDF stream processing engine for optimizing the logical query performances by the state of the streaming data. Strider is a framework often used for implementing adaptive properties such as high throughput,acceptable latency, and scalability. These frameworks are designed on state-of-the-art engine architectures using Apache Kafka and Apache Spark.

Ed-daoudy et al.[19] proposed a novel approach for detecting real-time health disease prediction using medical data streams with the help of Twitter data,Apache Kafka, and Apache Cassandra. They ingested the collected Twitter data using Spark streaming, and further machine learning algorithms were used for selected features. The experimental result shows an accuracy of 92.05% when using the Random Forest (RF) algorithm, which is quite high when compared to others.

Pereira et al. [20] presented a smart geo layer (SGeoL), an application for designing smart city applications. They integrated the geospatial data with urban data using an underlying middleware infrastructure for heterogeneous data processing and data context management. The experiment gives an empirical and scalable result for its performance.

**2.5 SPARQL for Complex event processing**

Cardinale et al. [21] highlighted the various flaws of an existing system for reducing the error related to the velocity and variety of data. The flaws include performance evaluation for RDF stream processing (RSP), working prototypes, and dependencies that increase the complexities, as well as benchmarking.

Mebrek et al. [22] introduced a novel framework for distributed computing using RDF stream processing (RSP) to accommodate heterogeneous and continuous data streams for instant processing. Due to the complexity of the queries, they found the need for multiple engines that can efficiently process queries. The primary experimental evaluation shows good results using multiple engines rather than existing ones.

## 3. Complex Event Processing for Intelligent Building Operations Monitoring and Prediction

Complex Event Processing (CEP) is a technique that is used to monitor and forecast intelligent building processes. It entails the real-time analysis and processing of massive amounts of data from numerous sources in order to spot patterns, detect complex events, and initiate relevant responses. CEP may be used to improve monitoring, prediction, and decision-making processes in the context of intelligent building operations. The fig 1 shows the proposed architecture for this research work.

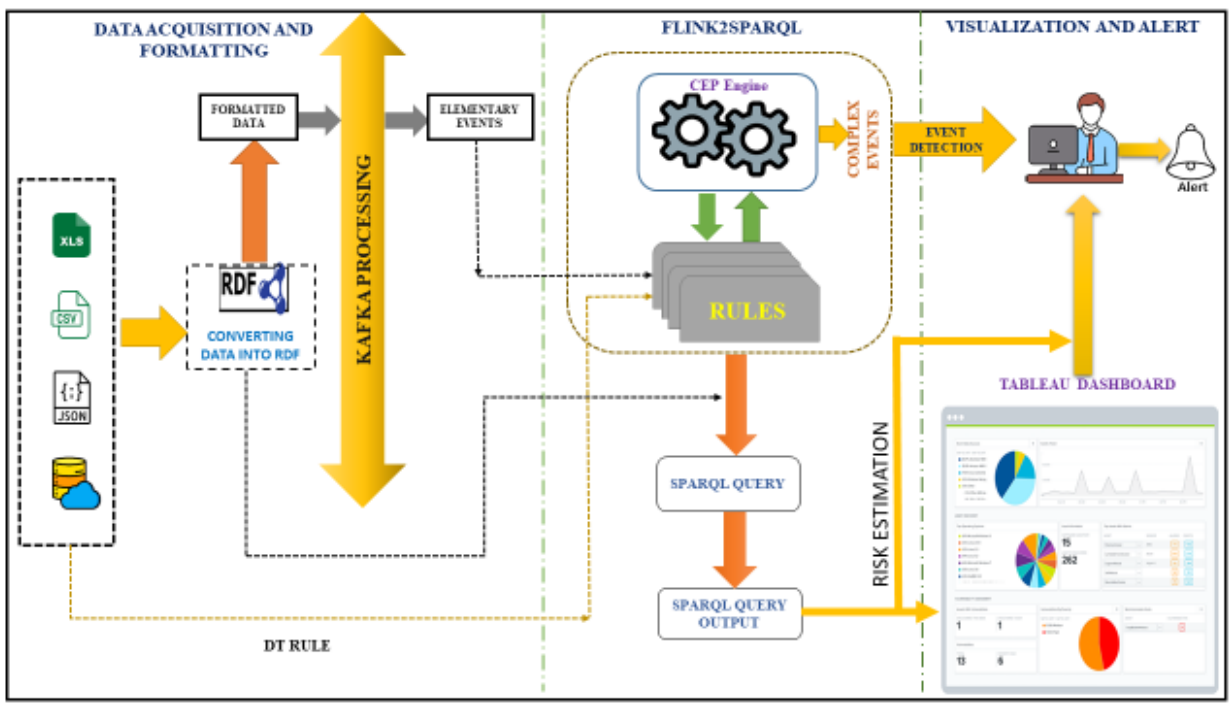

Fig:- 1 Proposed Architecture for CEPIBOP (Complex Event Processing for Intelligent Buildings Operations)

### 3.1 Dataset Description

In this section, we will elaborate on various features of the datasets. The datasets consist mainly of smart building operation features such as temperature, humidity, $CO_2$, energy, light status, etc. The format of the datasets is CSV. The dataset is further used for conversion into RDF datasets that allow for more effective decision-making using complex queries, which leads to more accurate discoveries and helps in more data-driven decision-making. The dataset is preprocessed before conversion into RDF data. The IoT dataset is collected from a repository that contains various features, including 8192 samples[1].

## 3.2 Resource Description Framework

The Resource Description Framework (RDF) is a generic framework for expressing web-connected data. RDF statements are used to describe and communicate metadata, allowing for the standardization of data sharing based on relationships. RDF is used to combine data from several sources. A website that offers online catalog listings from a manufacturer and links items to reviews on other websites and merchants selling the products is an example of this method. The semantic web is built on the RDF framework, which is used to organize information based on meanings. In the proposed architecture shown in Fig. 2, data is collected from various repositories. We use the GraphDB tool for the conversion of data into RDF data, which is in the triple format (subject-predicate-object). Onto Refine[2] is the plugin in GraphDB which helps in conversion of different types of format data into RDF[3] [23].

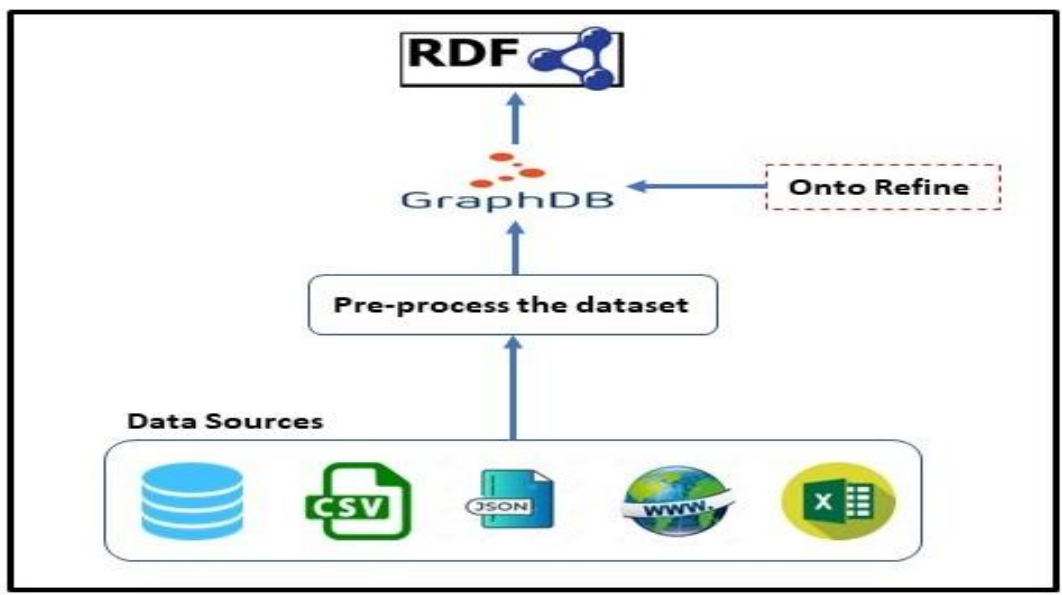

Fig:- 2 Datasets to RDF conversion

---

[1] https://www.kaggle.com/datasets/robmarkcole/occupancy-detection-data-set-uci
[2] https://www.ontotext.com/blog/tabular-data-rdf-graphdb/
[3] https://www.w3.org/RDF/

**3.3 RDF to Kafka processing**

This subsection elaborates on event flow from the RDF to stream event processing through Kafka producer and consumer methodologies. An RDF event producer first ingests the data through the publish and subscribe module, where the producer gathers and transmits the information to the internal processing through various brokers named A, B, and C, which subsequently carry out the scheduling and replication of events.

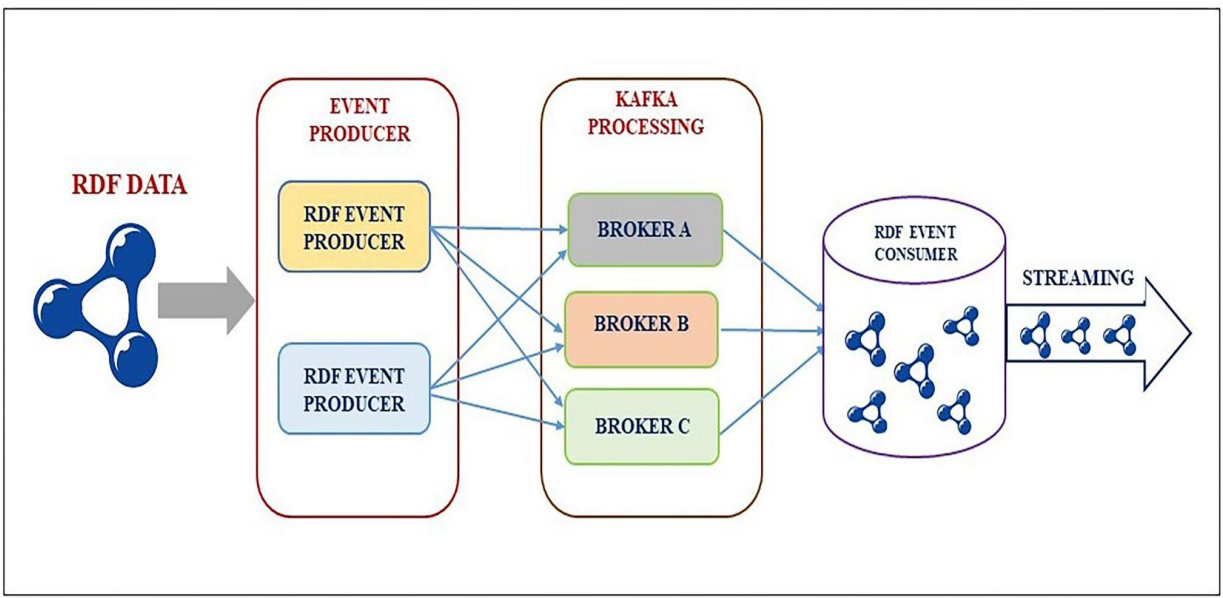

Fig:- 3 RDF Stream processing through kafka

Since streaming data is dynamic in nature and a predicted failure of the broker may lead to data loss, it is kept at multiple brokers. The RDF event consumer divides the incoming RDF stream into various events for later processing. Fig 3 depicts the producer and consumer for stream event processing through the Kafka framework. The fig 4 depicts the dataset ingestion into kafka for stream processing.

Fig:- 4  Stream processing with Kafka Producer Consumer

**3.4  CEP engine operation with Siddhi**

WSO2 Siddhi[4] is a complex event processing engine that listens to events from different data stream sources and determines the complex events based on a set of predefined rules. It is an open source analytics tool that allows deployment and builds applications that capture data from streams and detect patterns in real time [24]. It works with the SPARQL query language, in which users can set the predefined rules, and based on queries, action is taken. The siddhi CEP takes input in the form of streaming data, as shown in fig. 5. In this work, the streaming data in the form of RDF, which contains triples of data in the form of humidity, $CO_2$, temperature, occupancy, and light intensity, is passed to the rules of the CEP engine. The elementary rules are created using standard parameters and decision tree rules. The rules are analyzed using the Siddhi Core, and after analysis, the complex events are formed. Here, after analyzing the rules, the complex events are: humidity is high, temperature is moderate, air quality is poor, etc.

---

[4] https://siddhi.io/en/v5.1/docs/

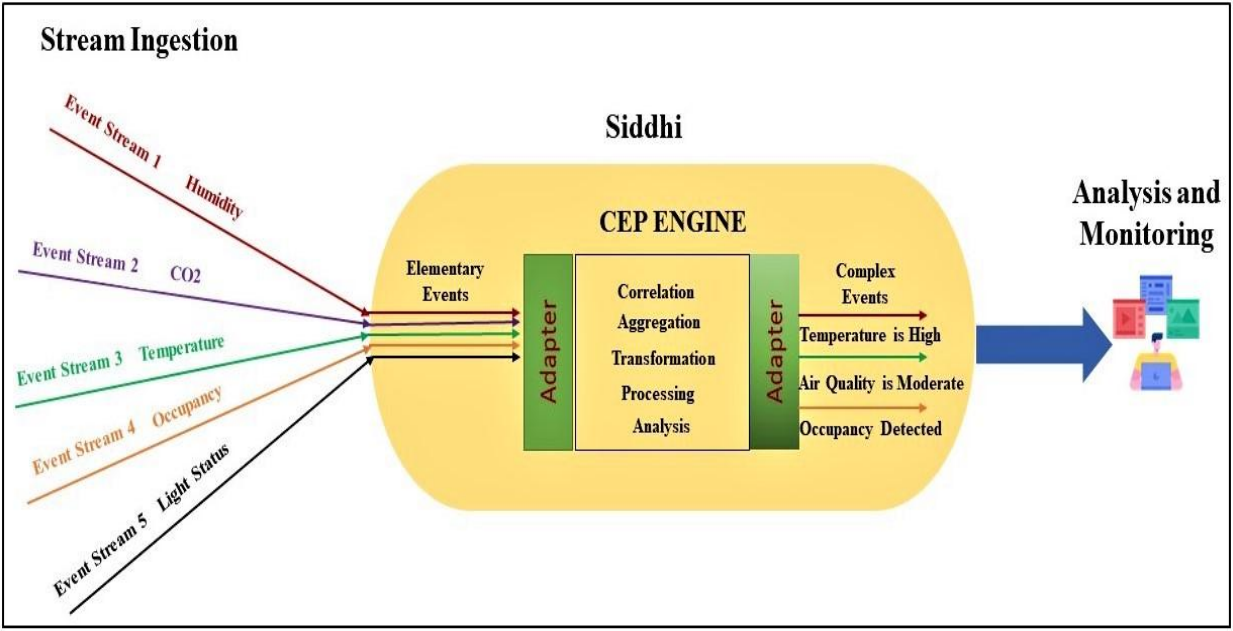

Fig: 5  Event stream processing through Siddhi CEP

**3.5 Flink to Sparql Connectivity**

Flink is a true streaming framework for efficiently managing various data sources within memory computation capabilities. The kernel of the framework is able to process the events consistently with marginal latency at very high velocity. This framework is an open-source distributed batch processing system. Regular programmes written in Java, Scala, or Python are essentially what Flink programmes are at their highest level. Flink programmes are translated to dataflow programmes, which perform a variety of transformations (such as filtering, mapping, joining, and grouping) on distributed collections that are first formed from other sources (e.g., by reading from files). In this work, the data flows through Kafka processing components and is passed through the Flink ecosystem, which is connected through rules that are generated through DT rules, and based on that, the query can be processed. The SPARQL query enables the Flink to SPARQL component to efficiently run the query and fetch the result in real time [26].

Table 1 : Pseudo code for proposed architecture

| **Steps for proposed architecture** |
|---|
| Require:-Different types of datasets.<br>Output:-Real time alert generation |
| **BEGIN**<br>STEP 1: Preprocessing of the dataset<br>      **If** (missing values are available)<br>      preprocess the data;<br>      **Else**<br>    b: Go to step 2<br>STEP 2: a:Use GraphDB and Onto Refine for Conversion into RDF data;<br>      b:Develop rules from the datasets using Decision Tree and<br>        also use standard parameters of Smart Building operations;<br>STEP 3: a:Stream RDF through Kafka;<br>      b. Move elementary events to Flink rules;<br>STEP 4: a:Ingest rules and RDF data and correlate using Siddhi CEP engine<br>       and detect complex events;<br>STEP 5: a:Generate Complex events with the help of rules;<br>STEP 6: a:Use SPARQL for query in real time based on events detected for better estimation of events;<br><br>STEP 7: a: Query on RDF data for extracting historical data;<br><br>STEP 8: a:For Risk estimation follow the results of STEP 6 and 7;<br><br>STEP 9: a: For Monitoring and Visualization use tableau for making this we get data with the help of SPARQL;<br>STEP 10: a: For ALERT Generation based on result of<br>      STEP 5(a), STEP 6, STEP 7 STEP 8 and STEP 9;<br>STEP 11: Return STEP 1;<br><br>**END** |

### 3.6 Rules development

In this subsection, we will elaborate on the rule generation terminology for this work. The rule generation is important for real-time event detection as well as based on event risk estimation or other operations that may be performed. The decision tree methodology is a commonly used data mining method for classification. It classifies the data in the form of an hierarchical tree, where (i) the root node defines the population of the various features. (ii) Decision nodes are formed by the splitting of root nodes. (iii) Branches represent the chances of outcome that emanate from the root and internal nodes. (iv) The node at which further splitting is not possible is called the leaf node of the decision tree. In fig. 1 rules are connected through the datasets used for this work, since one of the datasets contains occupancy as labeled data. For rule generation we have used a decision tree as it classifies the data based on labels as shown in the fig. 6. Based on the decision tree the rule is formed as shown in Table 2.

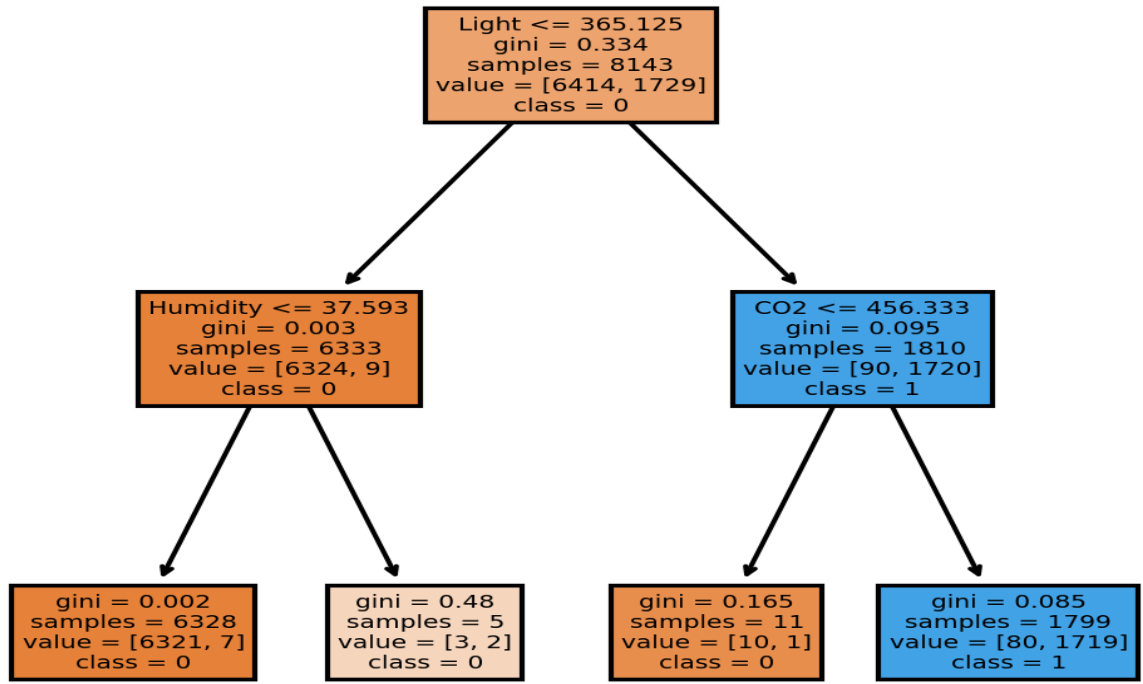

Fig: 6  Decision tree for rule formation 3.6.1 Rules

**Table:2** Rules formed using Decision Tree

| S.NO | Rules |
|---|---|
| Rule 1 | If light <= 365.125. then occupancy becomes false |
| Rule2 | If humidity <= 37.593 then occupancy becomes false |
| Rule3 | If light >= 365.128 then occupancy becomes true |
| Rule4 | If humidity>37.593 then occupancy becomes true |
| Rule5 | If $CO_2$ <= 456.333 then occupancy becomes true |
| Rule6 | If light <= 365.125 and light $CO_2$ <= 456.333 Then occupancy becomes true |
| Rule7 | If $CO_2$ <= 456.333 and light <= 365.125 |
|  | Then occupancy becomes true |
| Rule 8 | If humidity <37.593 and $CO_2$ <= 456.333 and light >= 365.125 Then occupancy becomes false. |

We have also developed various other rules related to closed premises temperature rise and gas rise detection that can be implemented inside smart buildings for any kind of risk situation, and appropriately, the decision can be made at the earliest possible time for evacuating the occupied premises. In this work, we have developed 37 rules for smart building premises that are helpful for developing early warning systems and keeping occupants informed during any emergency situation with real-time potential risks.

**3.7 SPARQL query working**

The SPARQL Query Language is a standardized language for querying RDF data and a protocol that enables the execution of SPARQL queries and the delivery of their answers via the Web. The RDF data model serves as the foundation for SPARQL [27], which is independent of RDF schema languages which is shown in fig. 7. In other words, it merely offers support for graph pattern matching for RDF graphs rather than any kind of reasoning skills. In terms of how relational databases are queried, it is comparable to the Structured Query Language (SQL).

**A SPARQL query typically consists of the five components listed below.**

1. IRI prefixes can be defined in Prefix declarations, so they can be utilized as shortcuts later in the query.
2. The dataset Clause feature enables the specification of a closed dataset partition over which the query should be conducted.
3. The result clause feature enables you to describe the kind of SPARQL query that is being used and, if applicable, the results that should be provided.
4. The query clause enables the specification of the query patterns that are used to construct the variable bindings for the defined variables in the query and to compare against the data.
5. Results can be ordered, divided, and paginated using solution modifiers.

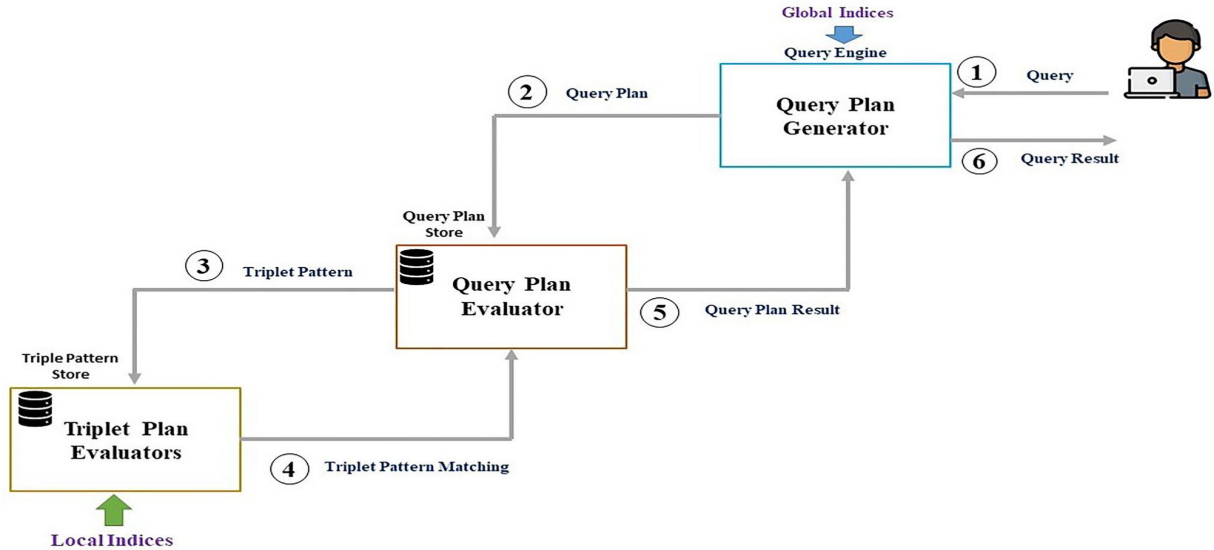

Fig: 7 SPARQL Query processing steps

### 3.7.1 SPARQL query results of different conditions using GraphDB

In fig. 8, it is shown that the framework view of the GraphDB tool for the query result, for which the date has a humidity greater than 30. Table 3 shows the same SPARQL query.

Fig:8 Query result for humidity

Table 3. SPARQL Query for Humidity

```
PREFIX ns1: <http://schema.org/>
PREFIX rdf: <http://www.w3.org/1999/02/22-rdf-syntax-ns#>
PREFIX xsd: <http://www.w3.org/2001/XMLSchema#>
SELECT ?date ?humidity
WHERE {
  ?description ns1:date ?date .
  ?description ns1:Humidity ?humidity .
  FILTER (xsd:float(?humidity) > 30)
```

In fig. 9, it is shown in the framework view of the GraphDB tool for the query result for which the date has a temperature greater than 30. Table 4 shows the same SPARQL query.

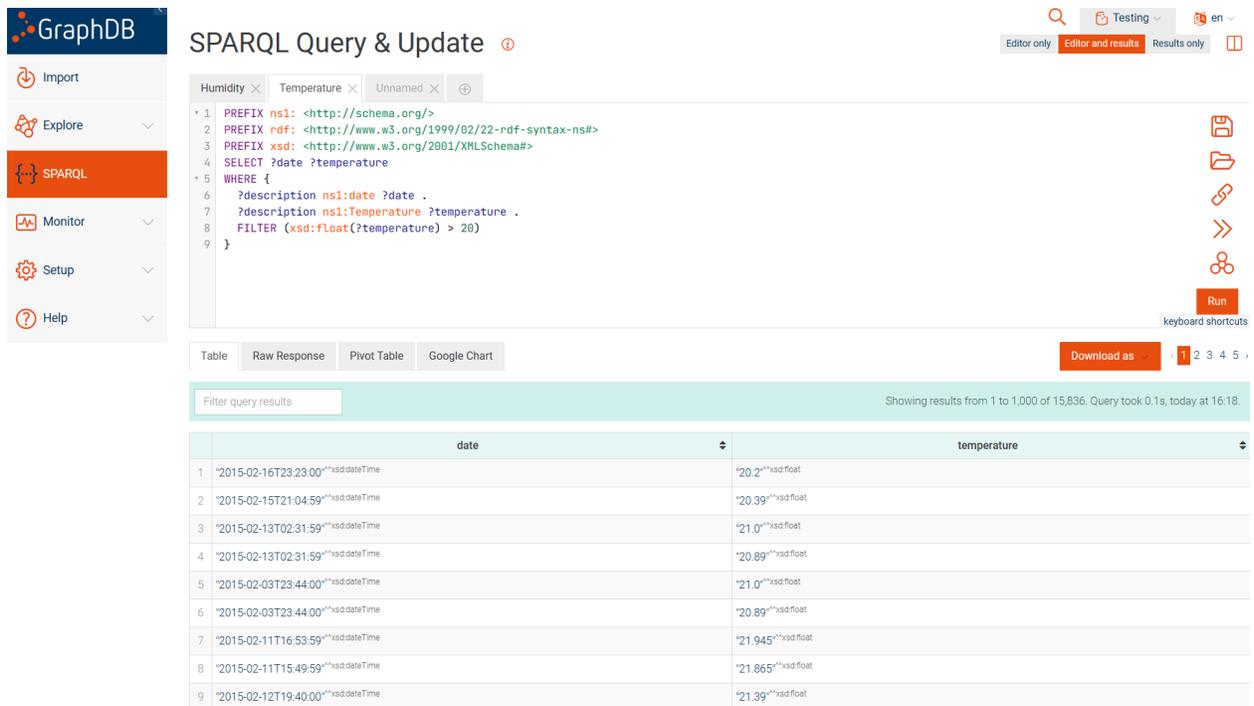

Fig:9 Query result for temperature

Table 4. SPARQL Query for Temperature

```
PREFIX ns1: <http://schema.org/>
PREFIX rdf: <http://www.w3.org/1999/02/22-rdf-syntax-ns#>
PREFIX xsd: <http://www.w3.org/2001/XMLSchema#>
SELECT ?date ?temperature
WHERE {
  ?description ns1:date ?date .
  ?description ns1:Temperature ?temperature .
  FILTER (xsd:float(?temperature) > 20)
}
```

In fig. 10, it is shown in the framework view of the GraphDB tool for the query results of all temperature details of premises between 1:30 and 2:00 PM. Table 5 shows the same SPARQL query.

Fig:10 Query result of temperature between 1:30 to 2:30 PM

Table 5. SPARQL Query for Temperature of time duration of 1:30 to 2:00 PM.

```
PREFIX ns1: <http://schema.org/>
PREFIX rdf: <http://www.w3.org/1999/02/22-rdf-syntax-ns#>
PREFIX xsd: <http://www.w3.org/2001/XMLSchema#>
SELECT ?date ?temperature
WHERE {?description ns1:date ?date .
  ?description ns1:Temperature ?temperature .
  FILTER (xsd:dateTime(?date) >= xsd:dateTime("2015-02-03T08:00:00") &&
xsd:dateTime(?date) <= xsd:dateTime("2015-02-04T23:00:00")
&&(xsd:float(?temperature) > 23))
```

According to the query in Table 6, we get whether the room is occupied or not, the $CO_2$ label of the room, and the humidity of the room, which is greater than 12. The result of the same query shown in fig. 11.

Fig:11 Complex query of $CO_2$, humidity and occupancy

Table 6. SPARQL Query for occupancy, $CO_2$ and humidity

```
PREFIX ns1: <http://schema.org/>
PREFIX rdf: <http://www.w3.org/1999/02/22-rdf-syntax-ns#>
PREFIX xsd: <http://www.w3.org/2001/XMLSchema#>
SELECT ?date ?CO2 ?occupancy ?humidity
```

> WHERE { ?description ns1:date ?date .
>  ?description ns1:$CO_2$ ?$CO_2$ .
>  ?description ns1:Occupancy ?occupancy .
>  ?description ns1:Humidity ?humidity .
>  FILTER (xsd:float(?$CO_2$) > 1100 && xsd:integer(?occupancy) = 1 && xsd:float(?humidity) > 12)}

### 3.7.2 Risk estimation through SPARQL

Risk estimation is based on two parts: the flink query result and the RDF query result. If an event is detected, determine whether it is normal, moderate, or risky. If the event is normal, there is no need to investigate further; if it is moderate, we go for the RDF query result and estimate what action to take, but the riskiest situation is handled easily. If the event is detected and its risk assessed, then query the RDF data and accordingly take instant action. We query on RDF data because we save all of our data in RDF format, so we need to query on an RDF dataset to get previous data. The fig. 12 shows about the risk estimation flowchart and how it is working. The risk estimation pseudo code is shown in table 7.

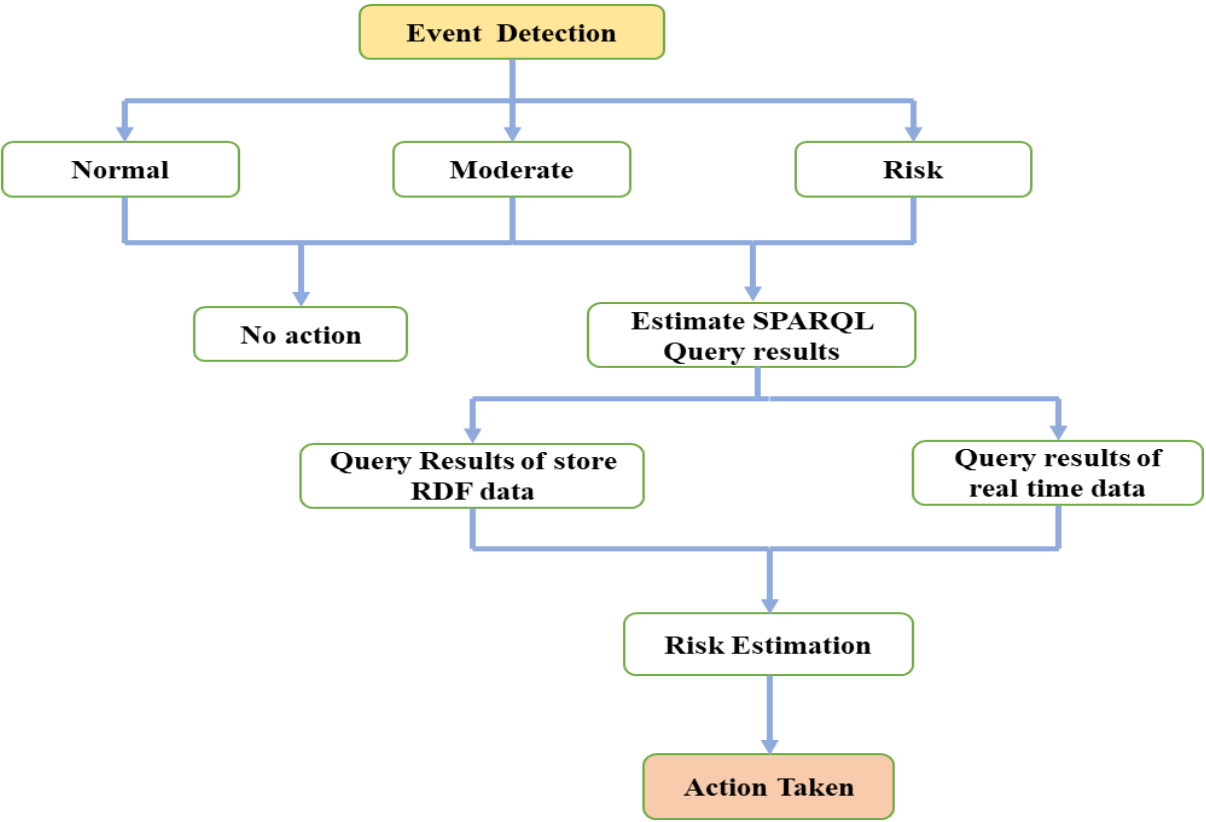

Fig:12 Risk estimation flow chart

**Table 7: Risk Estimation Pseudo code**

| Risk Estimation Pseudo code |
|---|
| **Require** : Input : T1 : event; T2 : value; T3 : for checking Normal, Moderate, Risk; T4: SPARQL query result based on RDF store data; T5: SPARQL query result on real time. T6: Alert given. |
| **For** <br> Event Detect <br> Check T1                                                     -> yes/no <br> Check T2                                              -> check event value <br> Check T3                                           -> Different risk parameter <br><br> **Goto** <br> T3 : Normal <br> Check T5: No Risk                                 -> Don't have to worry <br><br> **Goto** <br> T3: Moderate <br> Check T5                                   -> Chance of worry but not sure <br><br> **Else** <br> Check T4                                     -> Don't have to worry but take action <br><br> **Goto** <br> T3: Risk <br> Check T5                             -> High risk result instant action needed <br> Check T4                             -> RDF result is also supports of high risk <br><br> **Goto** <br> T6: Alert given <br><br> **Return   T1**                                    -> Check for new events |

### 3.7.3 Dashboard making

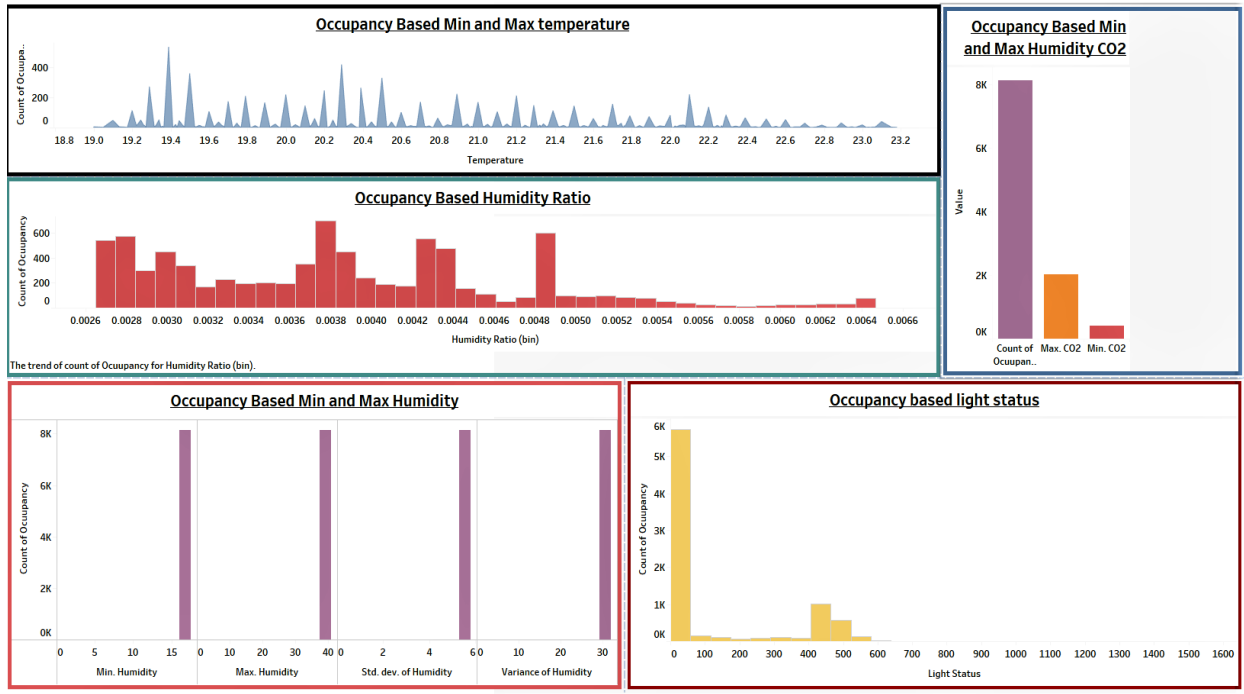

Fig:13  A demonstration of tableau dashboard for occupancy monitoring

According to the SPARQL query, we get the result, which has no need to be processed, and directly use this data in Tableau for making different worksheets that are combined into one main page, which is called the Dashboard page. On the main dashboard page, we have used a filter for better visualization.

## 4  Results and discussion

This section emphasizes the findings of this research work and its impact on the community working in this domain. The work has been proposed in the context of smart buildings, their operations, and how efficiently events are processed and important information is extracted through query processing.

To evaluate the performance of our CEP approach, we use a computer that has 16.0 GB of RAM, a 64-bit operating system, an x64-based processor, Windows 10 Pro Edition, and version 21H2 were employed as the hardware.

### 4.1 Performance of Event Processing Systems

We compare the performance with four different types of queries that were used in the process of detecting occupancy based on temperature, humidity, $CO_2$, and humidity ratio. For each case, 10,000 to 50,000 events were sent through the system, and the throughput of processing those events was measured. Table 8 shows how the query performs on different events in seconds. Query 1 is complex enough to require high performance in all four queries. Fig. 14 shows the number of events processed through simple and complex queries.

**Table 8:** The performance comparison of four diffirent types of query based on different events

| Query/Events | 10000 | 20000 | 30000 | 40000 | 50000 |
|---|---|---|---|---|---|
| Q1* | 879 | 768 | 701 | 611 | 506 |
| Q2 | 841 | 747 | 687 | 592 | 487 |
| Q3 | 821 | 731 | 663 | 578 | 456 |
| Q4 | 798 | 714 | 621 | 556 | 432 |

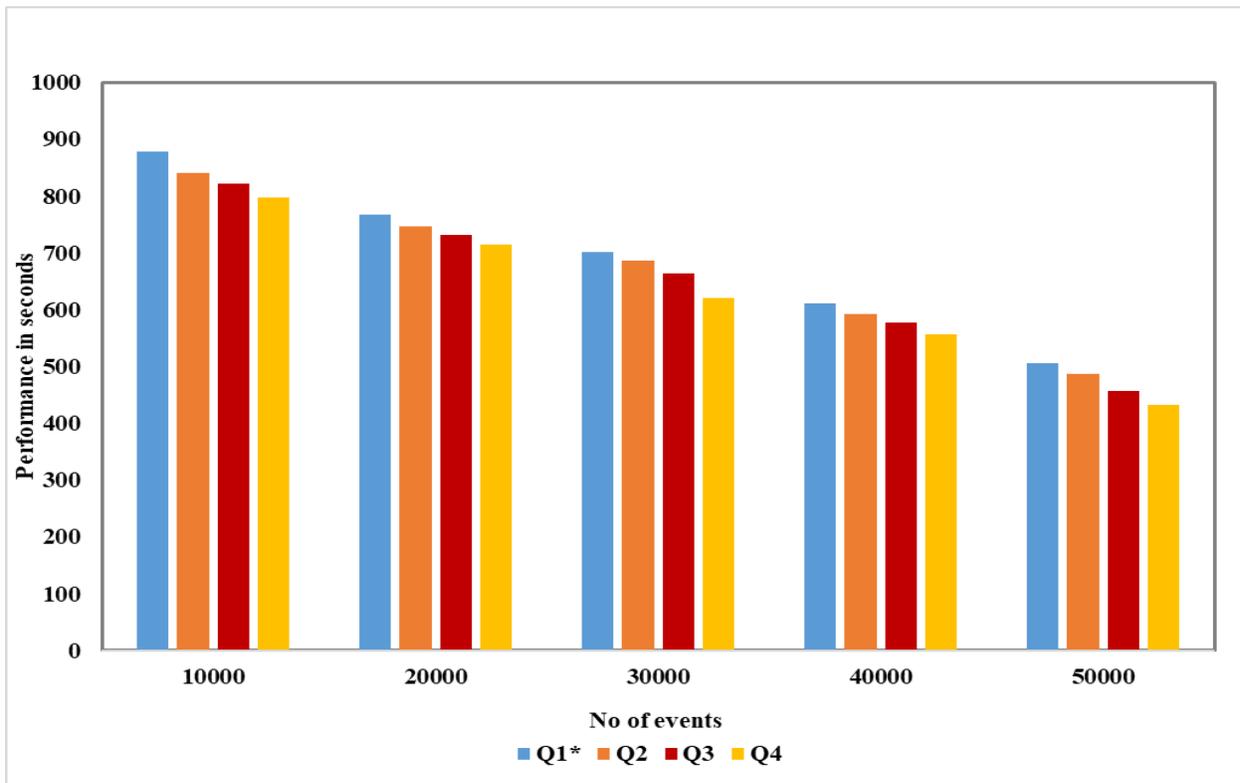

Fig 14 : Comparison of events for different queries

### 4.2 Rule Deployment Time

In this experiment, we evaluated the amount of time taken in the deployment of the rule in the CEP engine after making updates (e.g., inject, update, delete) on rules at runtime. The evaluation was conducted for two different cases: Without processing, in this case, the CEP engine is idle or No events. With processing: in this case, the CEP engine is processing different events (from 10,000 to 50,000 events per second). Each case was tested 20 times (i.e., updating rules by increasing their number each time) in order to calculate the average amount of time needed to deploy rules. Table 9 depicts the results of the experiments conducted for a single rule of different events. We conclude that the time(in seconds) of deployment increases when the number of events increases.

**Table 9:** The average time of rule deployment in different events.

| Rules /Events | 0 (Idle) | 10000 | 20000 | 30000 | 40000 | 50000 |
|---|---|---|---|---|---|---|
| R1 | 3.64 | 4.72 | 5.66 | 6.54 | 7.71 | 8.46 |

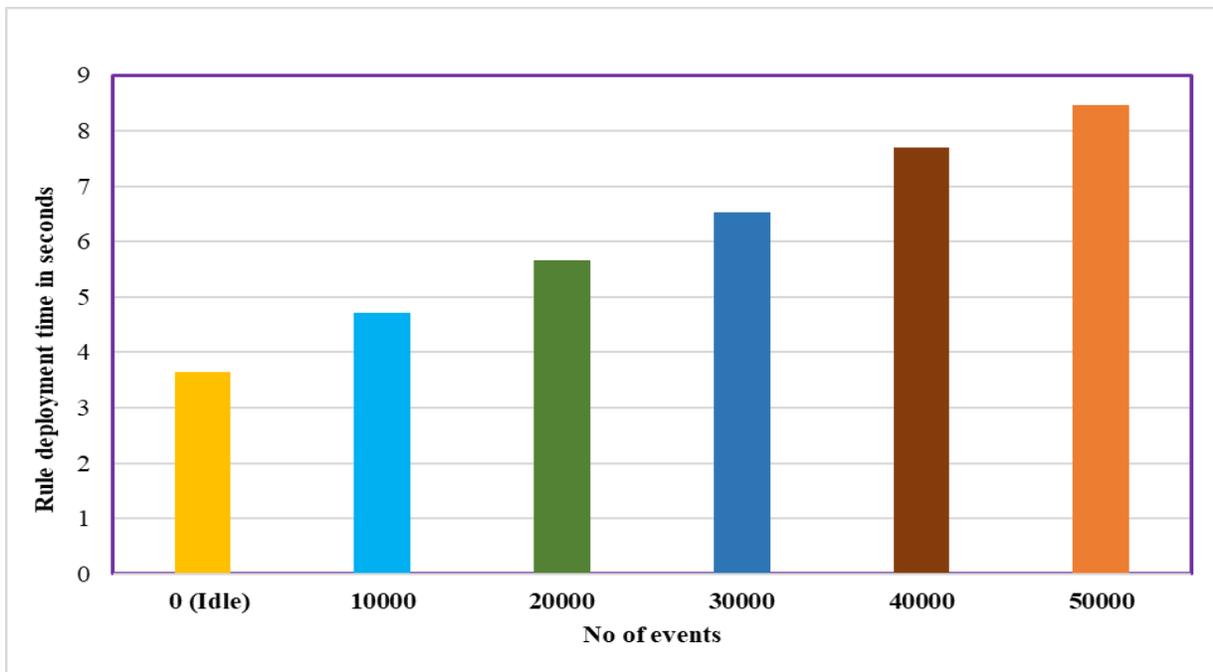

Fig 15: The average time of rule deployment

## 4.3 Calculation of Precision, Recall and F1 Score

In the context of smart building for evaluation of the proposed model it is important to detect the number of events out of all the events that occurred. We can evaluate the effectiveness of an event detection system in a smart building setting by measuring precision and recall and acquire insights into its accuracy and completeness in identifying events. For evaluating our proposed system we evaluate our model based on testing of 1000 events to calculate precision, recall, F1 score and accuracy as shown in table 10.

**Table 10**: Evaluation of metrics for event tested

| No of events tested | Accuracy | Precision | Recall | F1 Score |
|---|---|---|---|---|
| 1000 | 96.04% | 98.47% | 95.23% | 94.75% |

### 4.4 SPARQL query performance time based on RDF dataset

We have applied many queries to the RDF dataset [28], which makes the query performance time different on different datasets. The description of the different RDF datasets is shown in Table 11.

**Table 11:** Different RDF dataset description table

| Dataset | No. of triples | Classes | Entity | Size |
|---|---|---|---|---|
| RDF 1 | 8000 | 19 | 2200 | 1 MB |
| RDF 2 | 25000 | 19 | 3700 | 3MB |
| RDF 3 | 175000 | 19 | 30000 | 50 MB |
| RDF 4 | 1000000 | 32 | 376000 | 1 GB |
| RDF 5 | 2000000 | 48 | 460000 | 8 GB |

We have applied a total of six different queries to this dataset, and query execution time is affected by the size of the data and query complexity. All six queries have different complexity, which means that the query is different according to the event. The query performance based on time is shown in Table 12. The graph of the query performance time in seconds on different RDF datasets based on a single query is shown in Fig. 16.

**Table 12:** Show the query performance time in seconds on different RDF dataset.

| Query | RDF Dataset 1 | RDF Dataset 2 | RDF Dataset 3 | RDF Dataset 4 | RDF Dataset 5 |
|---|---|---|---|---|---|
| Q1 | 4 | 6 | 43 | 98 | 178 |
| Q2 | 8 | 12 | 56 | 119 | 212 |
| Q3 | 13 | 15 | 58 | 123 | 230 |
| **Q4*** | 36 | 86 | 121 | 174 | 314 |
| **Q5** | 21 | 33 | 46 | 67 | 129 |
| **Q6*** | 87 | 121 | 276 | 412 | 489 |

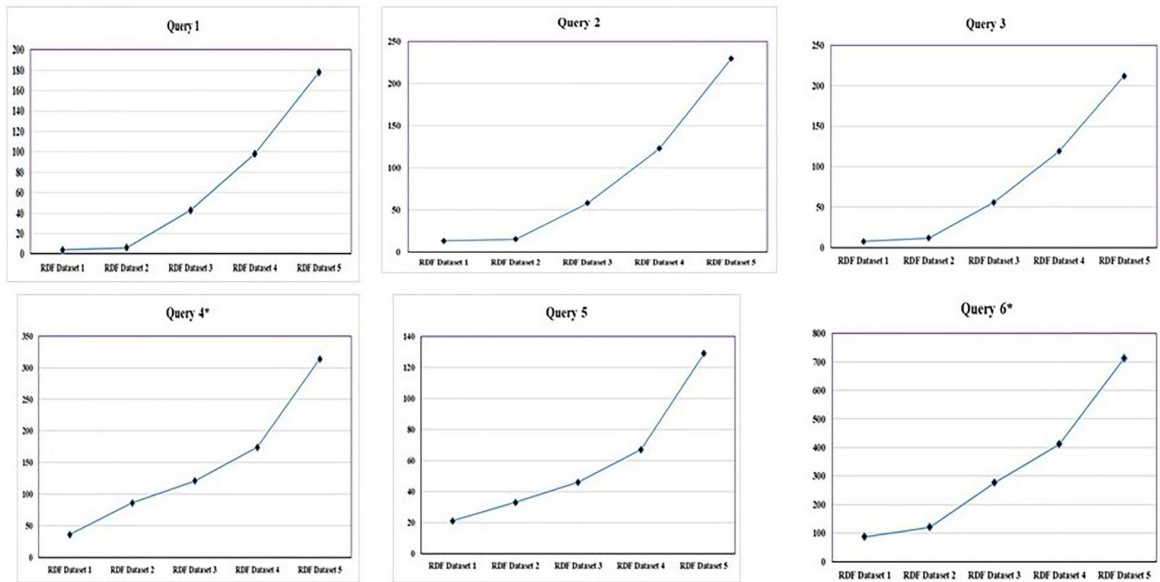

Fig 16: Query result of different RDF datasets.

Fig. 17 shows the query performance time based on different datasets.

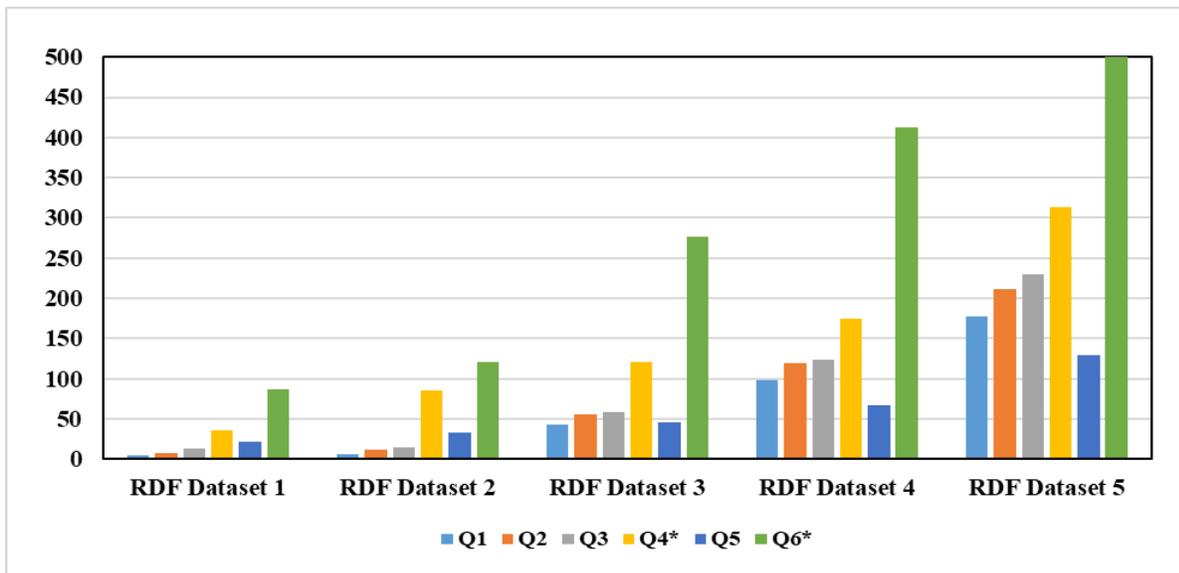

Fig 17 : Consolidated performance of queries on different datasets

When a query is executed, the CPU load on the system is high. Since the entire experiment was carried out on a good standalone system, the performance of the system while executing the simple and complex queries in terms of CPU load is shown in the fig. 18. The CPU load shown may increase as per the deployment based on events.

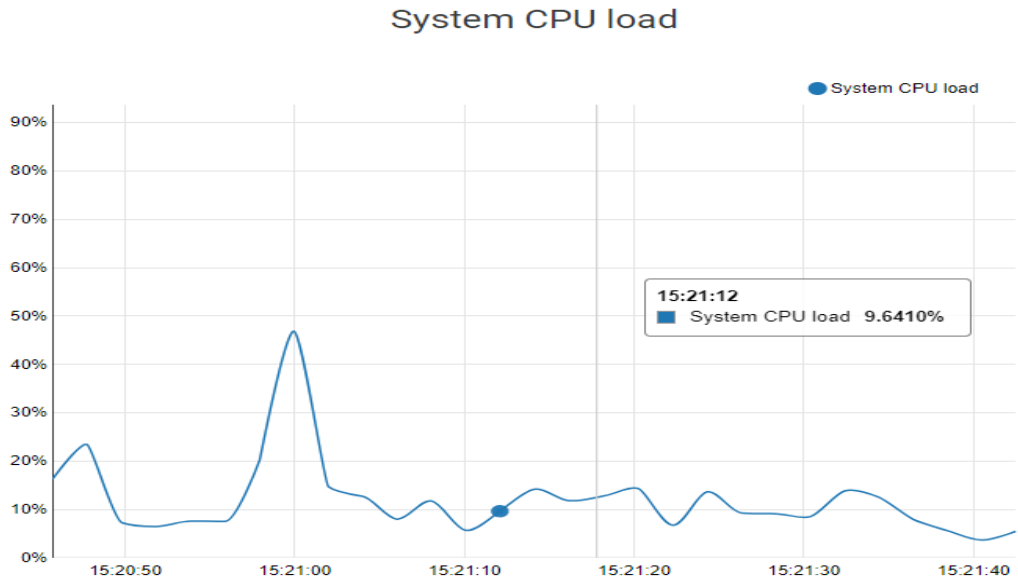

Fig 18 :- CPU performance during query processing

## 5. Conclusion and future scope

Smart buildings have emerged as a pioneering area of research for the IoT and are evolving continuously to fulfill the comfort needs of occupants to a great extent. In this context, we have proposed a complex event processing-based occupancy detection system to detect smart building operations in real time. In this context, we have converted IoT data into RDF with the help of the GraphDB tool. For streaming purposes, we have used Apache Kafka. Then we used Siddhi as a CEP engine to test the event streams in real time using a predefined set of rules extracted from Decision tree machine learning algorithms. Using these rules for event detection on stream data For risk estimation based on events, we used SPARQL queries for retrieving the information in real time and storing the data. The CEP engine is tested by deploying various rules on different events and testing query performance on different RDF datasets as well as different events. The events are tested with respect to evaluation metrics using precision, recall, and the F1 score. A number of complex queries are tested through SPARQL to check the performance of the proposed system that extracts useful information based on users' queries. Consequently, a risk estimation scenario is designed to alert the occupants inside smart buildings if any of the operations exceed the actual parameters so that an early warning system can be generated to evacuate the premises.

The future work may include finding robust models for rule generation as well as using a certified set of rules for designing smart building operations parameters, which will be more helpful for maintaining the occupants comfort inside a smart building.

**Acknowledgement**

This work is supported by the Department of Science and Technology, Government of India with File No. DST/ICPS/Individual-CPS/2018/T-244.